\documentclass[reprint,aps, prb, amsmath, amssymb, longbibliography,superscriptaddress]{revtex4-1}
\usepackage{amsmath}
\usepackage{bm}
\usepackage[normalem]{ulem}
\usepackage{graphicx}
\usepackage{braket}
\usepackage{float}
\restylefloat{table}
\usepackage[colorlinks, linkcolor= blue, citecolor = blue, urlcolor=blue]{hyperref}

\def\be{\begin{equation}}
\def\ee{\end{equation}}
\def \bea{\begin{eqnarray}}
\def \eea{\end{eqnarray}}
\def \nn{\nonumber}

\begin{document}
\title{Berry curvature induced nonlinear magnetoresistivity in two dimensional systems}
\author{Shibalik Lahiri}
\affiliation{Department of Physics, Indian Institute of Technology Kanpur, Kanpur 208016}
\author{Tanmay Bhore}
\affiliation{Department of Physics, Indian Institute of Science Education and Research Bhopal, Bhopal, 462066}
\author{Kamal Das}
\email{kamaldas@iitk.ac.in}
\author{Amit Agarwal}
\email{amitag@iitk.ac.in}
\affiliation{Department of Physics, Indian Institute of Technology Kanpur, Kanpur 208016}

\begin{abstract}
We investigate the effect of band geometric quantities on nonlinear magnetoresistivity, which dictates the quadratic dependence of the nonlinear voltage generated by the applied current. We propose that the interplay of the Berry curvature, the orbital magnetic moment and the Lorentz force can induce a finite nonlinear resistivity in two dimensional systems in presence of a perpendicular magnetic field. The induced nonlinear magnetoresistivity scales linearly with the magnetic field and is purely quantum mechanical in origin. Our proposed novel transport signature can be used as an additional experimental probe for the geometric quantities in intrinsically time reversal symmetric systems. 
\end{abstract}

\maketitle

\section{Introduction}

The band geometric properties of quantum materials, such as the Berry curvature and the orbital magnetic moment (OMM) play a fundamental role in the linear and nonlinear (NL) transport and optical properties~\cite{Xiao10, Gao19}. Some prominent examples, within the linear response, include phenomena like anomalous Hall effect (AHE)~\cite{Karplus54, sinitsyn_JPCM2008_semiclassical, Nagaosa_Ahe}, valley Hall effect~\cite{xiao_PRL2007_valley}, magnetic field induced -AHE~\cite{cullen_2021PRL_generating}, intrinsic Hall effect~\cite{gao_PRL2014_field, Das_PRB2021, tan_2021PRB_unconventional}, and magnetoresistance~\cite{zhou_2019PRB_valley,sekine_PRBR2018_valley}. Several very exciting NL Hall effects and other NL transport phenomena are also being actively explored~\cite{Deyo09, Moore10, Sodemann15, Xiao19, Kang19, du_NC2021_quantum, liu_arxiv2021_intrinsic,wang_arxiv2021_intrinsic,lai_NN_2021_third, liu_arxiv2021_berry}. However, the exploration of 
NL transport induced by geometric quantities in presence of a magnetic field is still at a nascent stage.

In presence of a magnetic field, the spin-orbit coupling has been shown to induce a unidirectional magnetoresistance in two dimensional (2D) magnetic systems~\cite{avci_NP2015_uni, yasuda_PRL2016_large,guillet_PRL2020_obser}. 
More recently, bilinear magnetoresistance~\cite{Zhang18, He18} and NL planar Hall effect~\cite{He19} were demonstrated in non-magnetic spin-orbit coupled 2D systems, based on the conversion of spin current to charge current. This was facilitated by including the magnetic field via Zeeman coupling. In this paper, we propose an alternative origin for NL magnetoresistance in 2D systems, which is purely quantum mechanical in nature. We show that band geometric quantities, such as the Berry curvature and the OMM induce second order NL magnetoresistance in presence of a {\textit {perpendicular}} magnetic field in 2D systems. This is even more significant in a time reversal symmetric system, where the entire contribution to the NL resistance arises from the band geometric properties. The predicted NL resistivity is a novel transport signature of band geometric properties and it can be used as an experimental tool to probe the Berry curvature and the OMM.

To obtain the NL resistivity, we use the semiclassical electron dynamics and the Boltzmann transport equation. 
We calculate all the NL conductivity contributions that arise from the Berry curvature, the OMM and the Lorentz force in 2D systems. We find that in addition to broken space inversion symmetry (SIS), the anisotropy of the band dispersion is also a necessary criteria for the NL conductivities to be finite. Intriguingly,
in contrast to the quadratic magnetic field dependence of the magnetoresistivity in the linear response regime~\cite{zhou_2019PRB_valley}, the predicted NL magnetoresistivity varies linearly with the magnetic field. Furthermore, it also persists in systems with broken time reversal symmetry (TRS), though there are additional contributions arising from the Drude conductivity. We explicitly calculate all the NL conductivities and the NL resistivities for 2D systems which host tilted massive Dirac fermions.

\begin{figure}[t]
\includegraphics[width=0.99\linewidth]{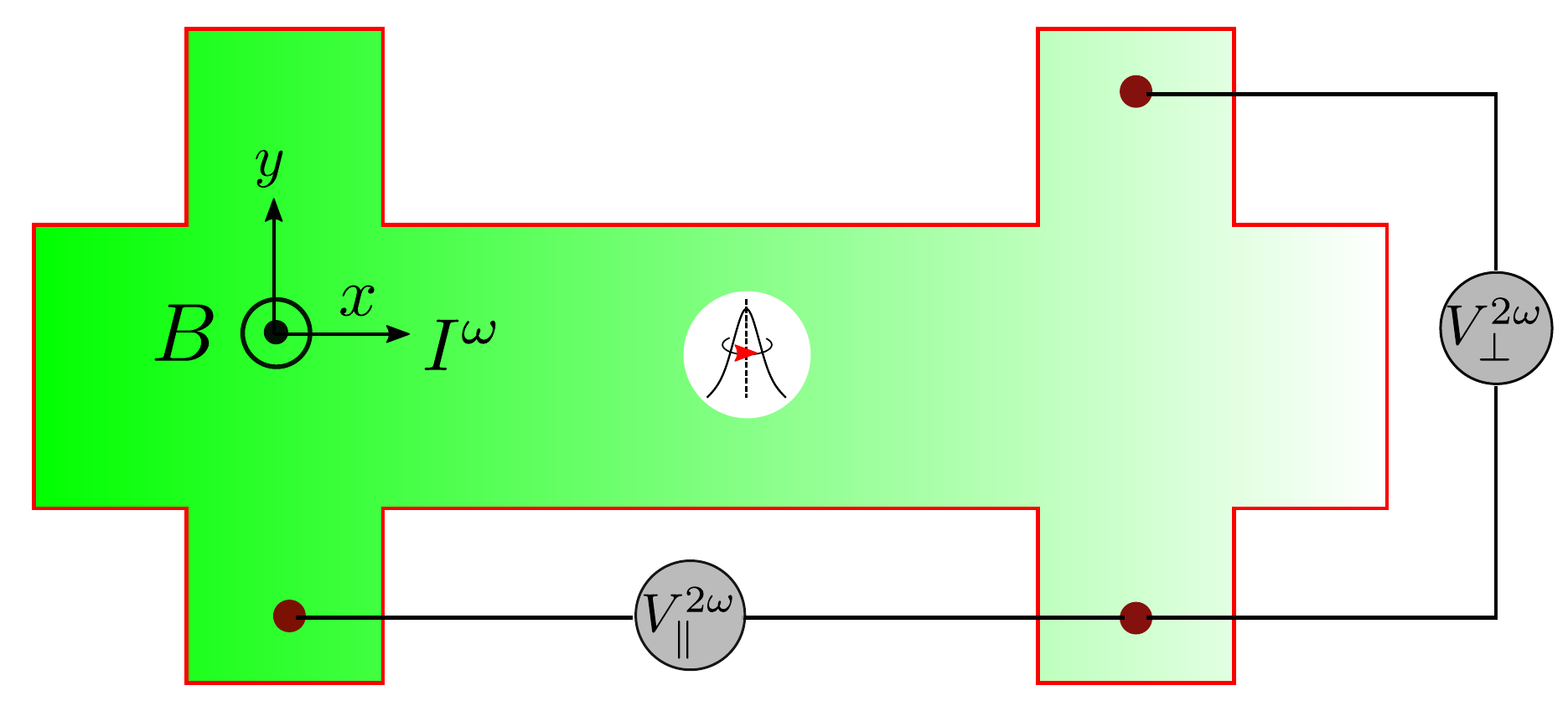}
\caption{Schematic of the experimental set up for the measurement of the NL magnetoresistivity. The magnetic field is applied perpendicular to the $x$-$y$ plane. The current (at frequency $\omega$) flows along the $x$-axis and the voltage along the direction of current - $V^{2\omega}_{\parallel}$, and perpendicular to it - $V^{2\omega}_{\perp}$, are measured.}
\label{fig_1}
\end{figure}

This paper is organized as follows: In Sec.~\ref{NL_R} we describe experimentally relevant NL resistivity matrix in terms of the theoretically calculated NL conductivities. This is followed by 
a detailed calculation of all the different contributions to the NL magnetoconductivities in Sec.~\ref{sbte}. 
Section~\ref{model_calc} presents a specific example of the predicted NL magnetoconductivity and magnetoresistivity in tilted massive Dirac Hamiltonian. This is followed by a discussion in Sec.~\ref{discussion} and finally we summarize our findings in Sec.~\ref{conclusion}.

\section{nonlinear resistivity}
\label{NL_R}

In this section, we will define the second order NL resistivity matrix and obtain its general expression in terms of the NL conductivities and the linear resistivities. Before we proceed to that, it is useful to understand the commonly followed experimental setup of the NL transport, shown in Fig.~\ref{fig_1}. In NL transport experiments~\cite{He18, Kang19, He19}, typically an ac current (or current density) of frequency $\omega$ is sent through the device, and as a response, the induced longitudinal~\cite{He18} (parallel to the current) and transverse~\cite{Kang19,He19} (perpendicular to the current) voltage drops (or electric fields) are measured. The induced NL voltages or electric fields, $E^{(2)}$, are distinguished from their linear counterparts by specifically measuring them at $2 \omega$ or $0$ frequency, using a lock-in amplifier. 

Specifically, for an input current density $j^\omega$, the induced linear response electric field $E^\omega$, and the induced NL responses in the form of the dc electric field $E^{\rm nldc}$, 
and the second harmonic electric field $E^{2\omega}$ are usually measured. 
This measurement scheme introduces the general concept of second order NL resistivity ($\tilde \rho^{(2)}_{abc}$) via the relation, $E^{(2)}_a = \tilde \rho^{(2)}_{abc} j_bj_c$, where $j_b$ is the applied current density. Here, the subscripts, $a/b$ denote the coordinate axes and summation over the repeated indices is implied. Particularly, we can define the NL dc (zero frequency) resistivity, $E^{\rm nldc}_a = \tilde \rho_{abb}^{\rm nldc} (j_b^\omega)^2$, and the second harmonic resistivity, $E^{(2\omega)}_a = \tilde \rho_{abb}^{(2 \omega)} (j_b^\omega)^2$. In addition to these, there will also be an induced electric field in the linear response, which defines the linear resistivity ($\rho_{ab}$), $E_a^{\omega} = \rho_{ab} j_b^\omega$.

To connect the experimentally measured NL resistivity to the theoretically calculated NL conductivity, it is essential to establish a well defined relation between the two. Theoretically, we calculate the second harmonic  current via $ j^{(2\omega)}_{a} = \sigma_{abc}^{(2 \omega)} E_b E_c$ and the NL dc current as $j^{\rm nldc}_{a} = \sigma_{abc}^{\rm nldc} E_b E_c^*$, where $E_b$ and $E_c$ denote the complex components of the applied electric field. 
One simple way to connect the NL conductivity with the NL resistivity, is to calculate ${j}_a^{\rm nldc}$ or $j_a^{(2\omega)}$ in terms of the components of $j_b^\omega$, making use of the linear response resistivity. This yields, $j_a^{2 \omega} = [\sigma^{2 \omega}_{ab'c'} \times \rho_{bb'} \times \rho_{cc'}]j_b^\omega j_c^\omega$. Now we can define the full resistivity as $\rho^{\rm total} \propto {E}/(j^{\omega} + j^{2\omega})$, and do a Taylor series expansion to obtain the second order NL resistivity. 
The details of the calculations are summarized in Appendix \ref{define_NL_R}.

Assuming that the current is applied only along the $x$ direction, for the case of 2D systems, the NL resistivity matrix elements are calculated to be
\be \label{NL_resistance}
\begin{pmatrix}
\tilde \rho_{xxx}^{(2)} \\
\tilde \rho_{yxx}^{(2)} 
\end{pmatrix}
= 
{-}[\rho] \cdot
\begin{pmatrix}
\sigma_{xxx} & \sigma_{xxy} & \sigma_{xyx} & \sigma_{xyy}  \\
\sigma_{yxx} & \sigma_{yxy} & \sigma_{yyx} & \sigma_{yyy} 
\end{pmatrix}
\begin{pmatrix}
\rho_{xx}^2 \\
\rho_{xx}\rho_{yx}\\
\rho_{xx}\rho_{yx}\\
\rho_{yx}^2
\end{pmatrix} .
\ee
Here, $\tilde \rho_{xxx}^{(2)}$ stands for the NL resistivity and $\tilde \rho_{yxx}^{(2)}$ stands for NL Hall resistivity. In Eq.~\eqref{NL_resistance}, the first term on the right hand side denotes the $2 \times 2$ linear response resistivity matrix for 2D systems. Note that Eq.~\eqref{NL_resistance} holds for both the NL dc resistivity as well as the second harmonic resistivity, depending on which NL conductivity is used on the right hand side of Eq.~\eqref{NL_resistance}.
However, in the transport limit, $\omega \tau \ll 1$, both the NL conductivities are identical [$\sigma^{\rm nldc} (\omega \to 0) = \sigma^{2\omega}(\omega \to 0)$], and this also reflects in the NL resistivity. 
For the rest of the paper, we work in this limit, and thus we have $\tilde{\rho}_{abc}^{(2)} = \tilde \rho_{abc}^{(2\omega)} = \tilde \rho_{abc}^{\rm nldc}$. Simplifying the notation further, we use 
$\tilde \rho_{xx}^{(2)}$ for $\tilde \rho_{xxx}^{(2)} $ and  $\tilde \rho_{yx}^{(2)}$ for $\tilde \rho_{yxx}^{(2)} $ in the rest of the paper.

To explore the NL magnetoresistivity in 2D systems induced by the Berry curvature and the OMM, we first calculate the different contributions to the NL conductivity. We will specifically consider the case of 
a magnetic field perpendicular to the 2D plane, {\it i.e.}, device geometry in the normal Hall configuration as shown in Fig.~\ref{fig_1}.

.

\section{Nonlinear conductivities} 
\label{sbte}

In this section, we calculate the general expressions of all the components of the NL conductivity tensor in presence of a magnetic field. As discussed in the last section, the NL conductivities are related to the NL current via the relation, $j^{(2\omega)}_{a} = \sigma_{abc}^{(2\omega)} E_b E_c$. In the semiclassical Boltzmann transport formalism, the charge current can be expressed as ${\bm j}(t)=-e \int [d{\bm k}] D^{-1}\dot {\bm r} g(t)$. Here, $g(t)$ denotes the non-equilibrium distribution function~(NDF), $[d{\bm k}]$ stands for $g_s d{\bm k}/(2\pi)^2$ with $g_s$ denoting the spin degeneracy. In a 2D system, in the presence of perpendicular magnetic field, the band geometric quantities modify i) the dynamics of the charge carriers in the phase-space ($\dot {\bm r}, \dot {\bm k}$), ii) the phase-space volume, $D^{-1}$ and iii) the band dispersion. These impact the NDF of the charge carriers in presence of applied electric field, which in turn gives rise to additional band geometry induced contributions to the NL conductivities. 

For the Hall configuration (${\bm E} \perp {\bm B}$) that we consider for this paper, the equation of motion is given by~\cite{Sundaram99, gao_PRL2014_field, Gao15}
\bea \label{band_velocity}
\dot {\bm  r} &=& D\left[\tilde{\bm v} + \dfrac{e}{\hbar}{\bm E}(t)\times  {\bm \Omega}\right],
\\
\hbar \dot {\bm  k} &=& D \left[-e {\bm E}(t) - e(\tilde{\bm v} \times {\bm B}) \right].
\eea
Here, $-e$ (with $e>0$) is the electronic charge and the phase-space modifying factor is given by $
1/D = \left[1 + \frac{e }{\hbar}({\bm B} \cdot {\bm \Omega}) \right]$ with ${\bm\Omega}$ as the Berry curvature. In the above equations, $\tilde {\bm v} = {\bm v} - {\bm v}_{\rm m}$ is the OMM modified velocity where $\hbar {\bm v} = \partial \epsilon/\partial{\bm k}$ and $\hbar {\bm v} _{\rm m}= \partial \epsilon_{\rm m}/\partial{\bm k}$ with $ \epsilon_{\rm m} = {\bm m} \cdot {\bm B}$. This is due to the fact that in presence of a magnetic field, the Zeeman like coupling of the OMM with the magnetic field modifies the electronic band energy via the relation, $\tilde \epsilon = \epsilon - {\bm m} \cdot {\bm B}$. 
The Berry curvature and the OMM for the $n$-th band can be computed using the relation~\cite{xiao_PRL2007_valley,Xiao07,Xiao10}
\bea \label{BC_formula}
\Omega_a^{n} &=& -2\varepsilon_{abc}\sum_{n\neq n'}\dfrac{{\rm Im}\bra{n}\partial \mathcal{H} /\partial k_b \ket{n'}\bra{n'}\partial\mathcal{H}/\partial k_c\ket{n}}{(\epsilon^n - \epsilon^{n'})^2},~~
\\\label{OMM_formula}
m_a^{n} &=& -\dfrac{e}{\hbar}\varepsilon_{abc}\sum_{n\neq n'}\dfrac{{\rm Im}\bra{n}\partial \mathcal{H} /\partial k_b \ket{n'}\bra{n'}\partial\mathcal{H}/\partial k_c\ket{n}}{\epsilon^n - \epsilon^{n'}}.~~
\eea
Here, $\varepsilon_{abc}$ is the Levi-civita symbol and the band energies and eigen-states are for an unperturbed system, $\mathcal{H} \ket{n}=\epsilon^n\ket{n}$ .
For 2D systems, both the Berry curvature and the OMM have only one finite component - the component pointing out of the plane. 
In our case, we consider the 2D system to be in the $x$-$y$ plane, thus only the $z$-components of these quantities are defined.

In presence of an ac electric field, of the form ${\bm E}(t) = {\bm E}e^{i\omega t} +  {\bm E}^*e^{-i\omega t}$, and a static magnetic field~\cite{Zyuzin18, Ma15} the NDF can be calculated using the Boltzmann kinetic equation with the relaxation time approximation~\cite{Ashcroft76}, which reads as
\begin{equation}\label{boltzmann}
\dfrac{\partial g (t)}{\partial t} + \dot{\bm k} \cdot {\bm \nabla}_{\bm k} g(t) =- \dfrac{g(t) - \tilde f}{\tau}.
\end{equation}
Here, $\tilde f$ is the Fermi-Dirac distribution function given by $\tilde f = 1/[1+e^{\beta(\tilde \epsilon - \mu)}]$ at chemical potential $\mu$ and inverse temperature $\beta=1/(k_B T)$ with $k_B $ being the Boltzmann constant and $T$ being the temperature. In Eq.~\eqref{boltzmann},  $\tau$ is the relaxation time and for simplicity we ignore its energy dependence. The NDF can be expressed as a sum of the equilibrium and non-equilibrium part, $g(t) = \tilde f + \delta g(t)$. Furthermore, the non-equilibrium part $\delta g(t)$ can be expressed as a power series of the applied electric field as $\delta g = \sum{_\nu} \delta g_\nu$ with $\delta g_\nu \propto |{\bm E}|^\nu $. In this paper we are interested in current $\propto |{\bm E}|^2$, and thus we calculate the NDF upto quadratic order in electric field. To this end, we use the ansatz,  
\be
\delta g_2 (t) =  f_2^{0} + f_2^{0*} + f_2^{2\omega} e^{i2\omega t} +  f_2^{2\omega *} e^{-i2\omega t}.
\ee
Here, the $f_2^0$ or $f_2^{0*}$ is the rectification (or dc) part and $f_2^{2\omega}$ or $f_2^{2\omega *}$ is the second harmonic ($2 \omega$) part of the NDF. Using this ansatz in Eq.~\eqref{boltzmann}, we calculate the second harmonic part to be~\cite{jones_PRSL1934_the,pal_PRB2010_nece}, 
\begin{equation}\label{dlta_n_2}
f_2^{2\omega} =\sum_{\nu=0}^\infty\left(D \tau_{2 \omega}\hat L_{\rm B}\right)^\nu D\dfrac{e\tau_{2\omega}}{\hbar}  {\bm E} \cdot {\bm \nabla}_{\bm k} f_1^{\omega}~.
\end{equation}
Here, $f_1^\omega$ is the linear order correction to the distribution function [see Appendix~\ref{NDF_linear} for details], and we have defined 
\be 
\hat L_{\rm B} =  \dfrac{e }{\hbar}(\tilde {\bm v} \times {\bm B})\cdot {\bm \nabla}_{\bm k} ~~~\mbox{and}~~~\tau_{2\omega} = \dfrac{\tau}{1 + i2 \omega \tau}.
\ee
The rectification counterpart can be obtained from Eq.~\eqref{dlta_n_2} simply by replacing $\tau_{2 \omega} \to \tau$ and ${\bm E} \to {\bm E}^*$.
Using Eq.~\eqref{dlta_n_2}, it is straightforward to calculate the  NDF as a power series of the magnetic field~\cite{Morimoto16a, Gao19, Zyuzin17a, Das19b,Das_PRB2021}. The explicit form of the distribution function is presented in the Appendix~\ref{NDF_nonlinear}. 

Using the obtained NL distribution function, one can calculate the rectification current ${\bm j}^{0}(t)$, the second harmonic current ${\bm j}^{2\omega}(t)$ and the corresponding conductivities. In this paper, we restrict ourselves to the lowest order magnetic field corrections to the NL conductivities. The details of the calculations and the general expressions for all the different conductivity terms are presented in Appendix~\ref{NDF_nonlinear}. Here, we list the second harmonic contributions and focus only on those contributions, which are non-zero in systems which intrinsically preserve TRS (non-magnetic systems). 

Before doing explicit calculations, we note that the scattering time dependence of the NL conductivities can be inferred from very general symmetry arguments. For example, if a current component $j \propto (B)^a (\tau)^b$, then under time reversal we have, $ -j \propto (-B)^a (-\tau)^b \propto  j\times (-1)^{a+b}$, and consequently $a+b$ should be an odd integer. Thus, terms with odd (even) powers of $B$ in the magnetoconductivity will always have even (odd) powers of $\tau$. This symmetry argument is also applicable for magnetic field independent NL conductivities, and in that case we consider $a=0$.

In intrinsically time reversal symmetric systems in 2D, we find that only the NL Hall conductivity~\cite{Sodemann15, Xiao19, Nandy19,du_NC2019_dis,du_NRP2021_non} survives in the absence of a magnetic field. It is given by
\be \label{sigma_NAH}
\sigma_{abc}^{\rm NAH} = - \dfrac{e^3 \tau_\omega}{2\hbar} \varepsilon_{abd}  \int[d{\bm k}]\Omega_d v_c  f' + \left(b \leftrightarrow c\right ).
\ee
Here, $f'\equiv \partial_\epsilon f$ is the derivative of Fermi function with respect to energy and we have defined $\tau_\omega\equiv\tau /(1 + i \omega \tau )$.
As the name suggests, this term contributes only to the Hall current and the diagonal components $\sigma_{aaa}^{\rm NAH}$ vanish. 
In presence of a magnetic field, we find that in addition to Eq.~\eqref{sigma_NAH} there are three other contributions to the NL conductivities: i) a contribution arising solely from the OMM ($\sigma^{\rm OMM}_{abc}$), ii) a contribution arising from the interplay of the anomalous velocity and the Lorentz force ($\sigma^{\rm AL}_{abc}$), and iii) a Berry curvature dependent contribution arising from the phase-space factor ($\sigma^{\rm B}_{abc}$). 

The OMM induced NL conductivity is given by
\bea \nn
\sigma_{abc}^{\rm OMM} &=&   \dfrac{ e^3\tau_\omega \tau_{2 \omega}}{2\hbar}  \int[d{\bm k}] \Big[v_{{\rm m} a}  \partial_{k_b} v_c  f' ~~~~~~~~~~~~~~~~~~
\\ \label{sigma_OMM}
& & + v_a \partial_{k_b} (v_{{\rm m} c} f' + \epsilon_{\rm m} v_c f'')\Big] + (b \leftrightarrow c).
\eea
Here, we have defined $f'' \equiv \partial_\epsilon^2 f$ and note that the derivative operator acts on all the terms appearing on its right side.
The anomalous velocity and the Lorentz force combine to give
\bea \nn
\sigma_{abc}^{\rm AL} &=& -  \dfrac{e^3\tau_\omega^2 }{2\hbar} \varepsilon_{abd}  \int[d{\bm k}]~\Omega_d \frac{eB}{\hbar}(v_y  \partial_{k_x}v_{c}  
\\ \label{sigma_AL}
& & -v_x  \partial_{k_y} v_{c} )  f' + (b \leftrightarrow c) .
\eea
This conductivity also contributes only to the Hall current. The NL conductivity contribution induced by the phase-space factor is given by
\be \label{sigma_B}
 \sigma_{abc}^{\rm B} = \dfrac{ e^3 \tau_\omega \tau_{2\omega}}{2\hbar} \int[d{\bm k}] v_a
(\Omega_{\rm B} \partial_{k_b} + \partial_{k_b} \Omega_{\rm B} ) v_c  f' + (b \leftrightarrow c).
\ee
Here, we have defined $\Omega_{\rm B} \equiv \frac{e}{\hbar} {\bm \Omega}\cdot {\bm B}$. It is clear from Eqs.~\eqref{sigma_NAH}-\eqref{sigma_B} that all these NL conductivities depend on either the Berry curvature or on the OMM. Based on this, we conclude that in intrisicaly TRS preserving systems, the second order NL responses are induced only by the geometric properties of the electron wave-function. 
In other words, in intrinsically TRS preserving systems, the Lorentz force by itself, without the Berry curvature or the OMM, cannot induce second order NL response in 2D systems.  

To calculate the NL resistivity, we also need the linear response conductivity matrix. 
The general expression of the linear response current, ${\bm j}^\omega(t)$, is calculated in Appendix~\ref{NDF_linear}. The Drude conductivity is given by $\sigma_{ij}^{\rm D} =-e^2 \tau_\omega \int [d{\bm k}] v_i v_j f'$. In an intrinsically TRS preserving system, the non-zero conductivity up to linear order in the magnetic field is given by~\cite{Hurd72, Ziman72, Das_PRB2021}
\bea \label{lorentz} 
\sigma_{ab}^{\rm L} &=& -e^2 \tau_\omega^2  \dfrac{eB}{\hbar}\int[d{\bm k}] v_a (v_y \partial_{k_x} v_b - v_x \partial_{k_y} v_b)  f'~,
\\\label{OMM_linear}
\sigma_{ab}^{\rm O} &=& \dfrac{e^2}{\hbar}  \varepsilon_{abd}  \int[d{\bm k}]  \epsilon_{\rm m} \Omega_d  f'~.
\eea
The superscript `${\rm L}$' implies the Lorentz force contribution (the normal Hall effect) and the superscript `${\rm O}$' denotes the OMM contribution (OMM induced Hall effect). 

This summarizes the general framework for calculating the second order responses. We now explicitly calculate the NL magnetoconductivity in quantum systems which can be described via pair of valleys hosting tilted massive Dirac cones. This is one of the simplest system, which has an anisotropic band dispersion along with finite band geometric quantities, and which can be treated analytically. We show explicitly in the next section that the anisotropy of the band dispersion is a necessary condition to obtain non-zero NL conductivities.


\begin{figure}[t]
\includegraphics[width=0.99\linewidth]{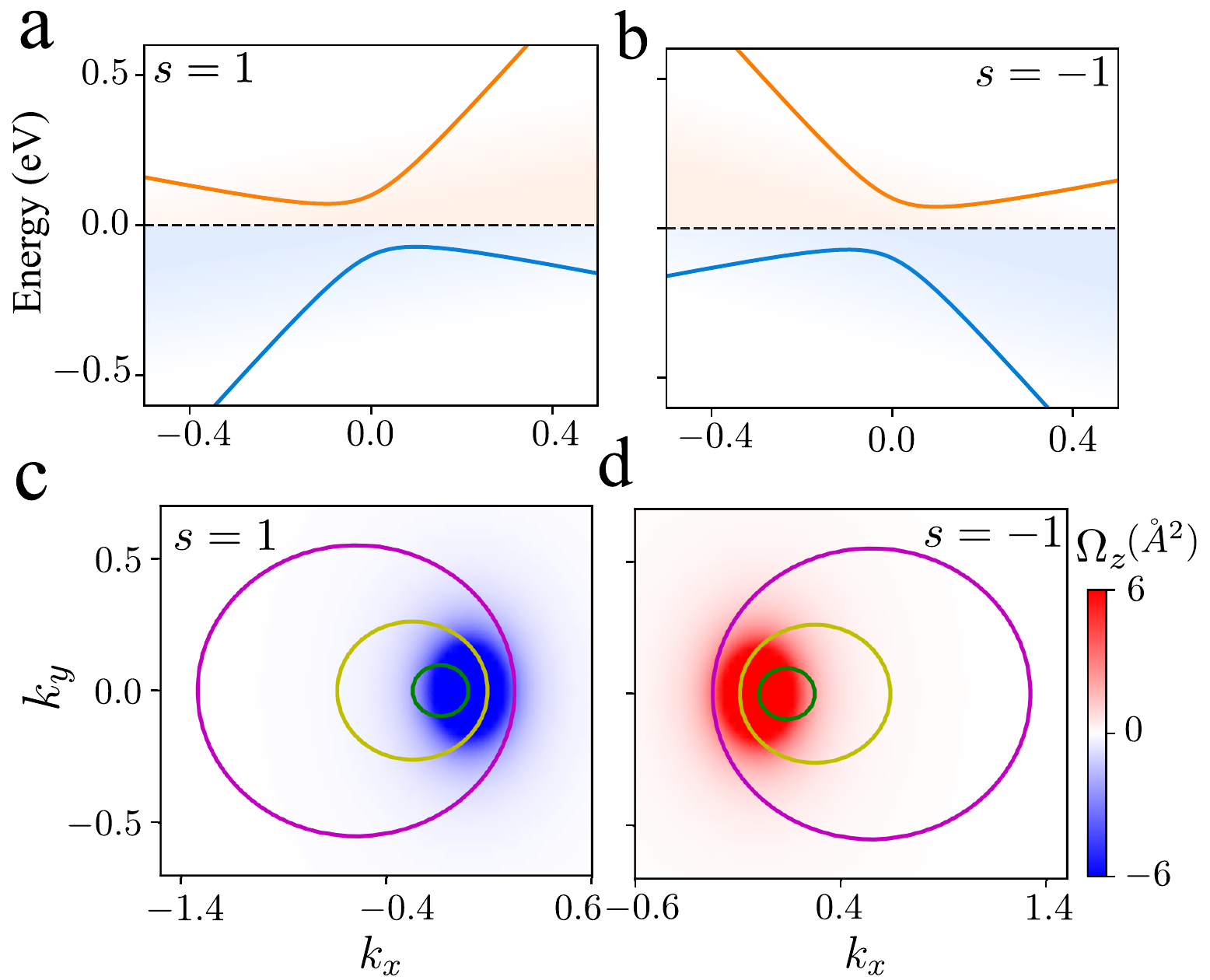}
\caption{The band dispersion of the tilted massive Dirac Hamiltonian along the $k_x$-axis for nodes with a) $s = 1$ and b) $s =-1$. Panels c) and d) shows the Berry curvature distribution in the momentum space for the conduction band for the nodes $s=1$ and $s=-1$, respectively. The different energy contours are indicated by the green ($\mu=0.1$ eV), yellow ($\mu=0.2$ eV) and magenta ($\mu=0.4$ eV) lines. The various parameters associated with the Hamiltonian are  $v_F =1$ eV$\cdot $\AA, $v_t=0.1 v_F$ and $\Delta = 0.1$ eV.}
\label{fig_2}
\end{figure}

\begin{figure}[t]
\includegraphics[width=0.99\linewidth]{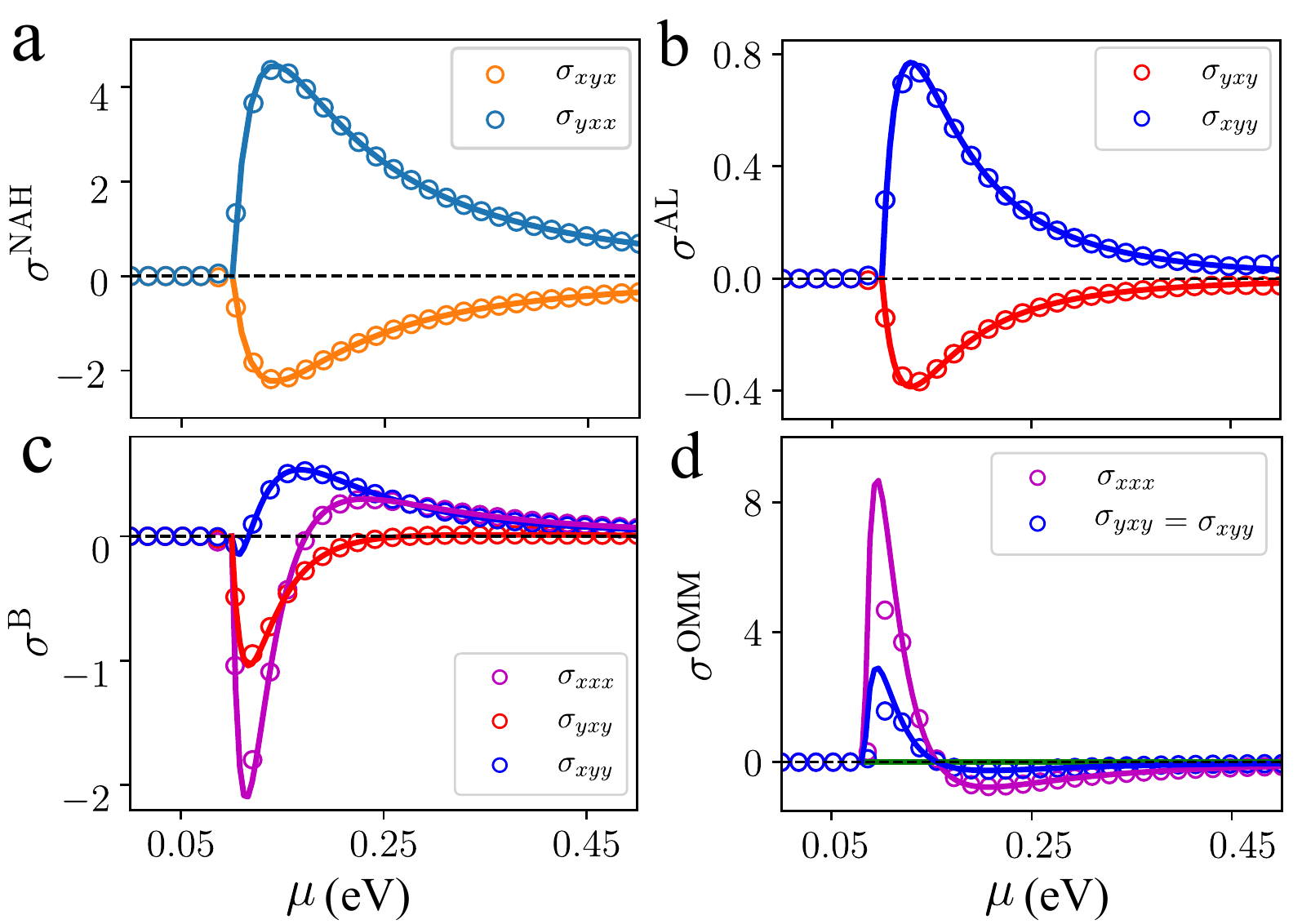}
\caption{Variation of the four different contributions to the NL conductivity with chemical potential~($\mu$). The open circles in the figures are results of numerical calculations at temperature $T=50$ K while the solid lines are the results of analytical calculations at zero temperature, upto linear order in $v_t$. a) The nonlinear anomalous Hall conductivity. b) The anomalous velocity and Lorentz force induced conductivity. c) The Berry curvature of phase-space factor induced conductivity. d) Orbital magnetic moment induced nonlinear conductivity. The conductivities are plotted in units of $10^{-3}$ nA.m/V$^2$. The Hamiltonian parameters are same as Fig.~\ref{fig_2}. Additionally we have considered $\tau=1$ ps. }
\label{fig_3}
\end{figure}

\begin{table*}[t!]
	\caption{Analytical results of the different contributions to the NL conductivities. For conciseness we have defined $r = \Delta/\mu$. The Berry 		
	curvature dipole, Lorentz force and anomalous velocity, phase-space factor and OMM induced NL conductivities are written 
	in units of $\{ \tilde{\sigma}^{\rm NAH}, \tilde{\sigma}^{\rm AL},\tilde{\sigma}
	^{\rm B} = \tilde{\sigma}^{\rm OMM}\} = g_s g_v \frac{e^3\tau_{\omega}}{\hbar^2}\frac{\Delta v_t}{4\pi\mu^2} \{1, \frac{e v_F^2 \tau_\omega}{\hbar^2 \mu} B, \frac{e v_F^2 \tau_{2\omega}}{\hbar^2 \mu} B\}$.
	}
	\begin{tabular}{cc ccc}
	\hline \hline
	~~{\rm NL
	} ~ & Anomalous Hall & Lor. force \& Berry curvature & Phase-space factor & Orbital magnetic moment 
	\\  conductivities~ &~~~~NAH [Eq.~\eqref{sigma_NAH}]~~~~& ~~~~AL [Eq.~\eqref{sigma_AL}] ~~~~ &~~~~ B 			
	[Eq.~\eqref{sigma_B}]~~~~  & ~~~~OMM [Eq.~\eqref{sigma_OMM}]~~~~
	\\ \hline \hline
	\rule{0pt}{3.5ex} 
	$\sigma_{xxx}$  & ~~$0$~~   & ~~$0$~~ & ~~$\tilde{\sigma}^{\rm B}\left(\frac{15}{4}  -15r^2 + \frac{45}{4}r^4\right) $~~ & 	~~$ -\tilde{\sigma}^{\rm OMM}\left(\frac{27}{4} - 21r^2 +\frac{45}{4}r^4\right)$~~
	\\[1.2ex]  
	\rule{0pt}{3.5ex} 
	$ \sigma_{yyx} = \sigma_{yxy}$   & ~~$0$~~  & ~~$-\tilde{\sigma}^{\rm AL} \frac{3}{4}\left(1-r^2\right)$~~   & ~~$			\tilde{\sigma}^{\rm B}\left(\frac{1}{2}  -\frac{17}{4}r^2 + \frac{15}{4}r^4\right)$~~  & ~~$-\tilde{\sigma}^{\rm OMM}		\left(\frac{9}{4}  -7r^2 +\frac{15}{4}r^4\right)$~~
	\\[1.2ex]  
	\rule{0pt}{3.5ex} 
	$ \sigma_{xyy} $  & ~~$0$~~  & ~~$ \tilde{\sigma}^{\rm AL} \frac{3}{2}\left(1-r^2\right)$~~  & ~~$\tilde{\sigma}^{\rm B}	\left(\frac{11}{4}  -\frac{13}{2}r^2 + \frac{15}{4}r^4\right)$~~  & ~~$-\tilde{\sigma}^{\rm OMM}\left(\frac{9}{4} - 7r^2 +	\frac{15}{4}r^4\right)$~~
	\\[1.2ex]  
	\rule{0pt}{3.5ex} 
	$ \sigma_{xyx} = \sigma_{xxy} $  & ~~$-\tilde{\sigma}^{\rm NAH}\frac{3}{4}\left(1 -r^2\right)$~~  & ~~$0$~~  & ~~$0$~~  & 	~~$0$~~
	\\[1.5ex]  
	\rule{0pt}{3.5ex} 
	$ \sigma_{yxx} $  & ~~$\tilde{\sigma}^{\rm NAH}\frac{3}{2}\left(1 -r^2\right)$~~  & ~~$0$~~  & ~~$0$~~  & ~~$0$~~
	\\[1.5ex]  \hline \hline
\end{tabular}
\label{table_1} 
\end{table*}

%
\section{tilted massive Dirac systems}
\label{model_calc}

In this section, we calculate the NL resistivity for 2D systems with a pair of tilted massive Dirac fermions. Each one of the `Dirac valley' is specified by the Hamiltonian~\cite{Sodemann15,du_PRL2018_band},
\be \label{ham}
{\mathcal H}_s = v_F  (k_x \sigma_y - sk_y\sigma_x) + s v_t k_x + \Delta\sigma_z~.
\ee
Here, $s=\pm$ is the valley index, $\sigma_i$'s are the Pauli matrices representing the sub-lattice degree of freedom, $\Delta$ is the band gap, $v_F$ denotes the Fermi velocity and the $v_t$ term introduces tilt in the band dispersion along the $k_x$-axis. The model Hamiltonian in Eq.~\eqref{ham} lacks SIS and the two valleys are related by the TRS. Furthermore, the mirror symmetry is broken along the $k_x$ line and preserved along the $k_y$ line~\cite{Nandy19}. This model
 acts as a building block of realistic band structures in systems like the surface states of topological crystalline insulators such as SnTe and transition metal dichalcogenides such as WTe$_2$~\cite{Sodemann15,du_PRL2018_band}. 
 
The energy dispersion for this two-band model is given by $\epsilon^{\pm} = sv_tk_x \pm \epsilon_0$, where $\epsilon_0 = (v_F^2 k^2 + \Delta^2)^{1/2} $ with $k = (k_x^2 + k_y^2)^{1/2}$. Here, the $+ (-)$ sign stands for the conduction (valence) band. The band dispersion for both the valleys is shown in Fig.~\ref{fig_2}(a)-(b). The tilt modified band velocity along the $x$-direction is given by $v_x^\pm = sv_t \pm v_F^2k_x/ \epsilon_0$, and $v_y^\pm =\pm v_F^2k_y/\epsilon_0$. 
For the model Hamiltonian in Eq.~\eqref{ham}, the Berry curvature and OMM have been calculated from Eq.~\eqref{BC_formula} and Eq.~\eqref{OMM_formula} respectively, and are given by~\cite{Sodemann15,du_PRL2018_band}
\bea\label{BC}
 \Omega_z &=& \mp s \dfrac{v_F^2\Delta}{2(v_F^2k^2+\Delta^2)^{3/2}}~,
\\\label{OMM}
 m_z &=&- s \dfrac{e v_F^2\Delta}{2\hbar(v_F^2k^2 +\Delta^2)}~.
\eea
As expected, the band edges are rich in both the Berry curvature and the OMM, and act as hotspots. The distribution of the Berry curvature with constant energy contours (in conduction band) are shown for both the valleys in Fig.~\ref{fig_2}(c)-(d). We note that both the OMM and the Berry curvature are independent of the tilt.

\subsection{Nonlinear conductivities}

Next, we calculate the NL conductivities for the tilted massive Dirac Hamiltonian in Eq.~\eqref{ham}, using the general Eqs.~\eqref{sigma_NAH}-\eqref{sigma_B} for the NL conductivities. For analytical insights, we calculate the NL conductivities upto leading order (linear order) in the tilt, and the results are summarized in Table~\ref{table_1} in terms of the parameter, $r\equiv \Delta/\mu $, for $\mu >\Delta$. We find that the NL conductivities are valley degenerate, thus we simply multiply the results for one valley with a valley degeneracy factor $g_v$. As a double-check of our calculations, we have also done numerical computation of the NL conductivities, including the tilt to all orders, and we find a reasonable agreement between the analytical and numerical results [see Fig.~\ref{fig_3}].
 
From the NL conductivities summarized in Table~\ref{table_1}, it is evident that the tilt in the dispersion plays an important role. More specifically, the tilt manifests in the anisotropy of the Fermi surface, and the $x$-component of the band velocity. 
Among the different components of the NL Hall conductivity, we find $\sigma^{\rm NAH}_{xyx} (=\sigma^{\rm NAH}_{xxy})$ and $\sigma_{yxx}^{\rm NAH}$ to be non-zero. This can be attributed to the non-zero Berry curvature dipole~\cite{Sodemann15} of the system which arises due to the broken mirror symmetry along the $x$-axis. The chemical potential dependence of $\sigma^{\rm NAH}_{xyx} (=\sigma^{\rm NAH}_{xxy})$ and $\sigma_{yxx}^{\rm NAH}$ is shown in Fig.~\ref{fig_3}(a). 
Similarly for the other Hall component which originates from the combined effect of anomalous velocity and Lorentz force, we find $\sigma^{\rm AL}_{yyx}(=\sigma_{yxy}^{\rm AL})$ and $\sigma^{\rm AL}_{xyy}$ to be non-zero. The chemical potential dependence of these terms is shown in Fig.~\ref{fig_3}(b). Unlike the above mentioned conductivity components, the phase-space contribution, shown in Fig.~\ref{fig_3}(c), and the OMM contribution, shown in Fig.~\ref{fig_3}(d), induced NL conductivities contribute both to the diagonal as well as to the Hall components. Both of these contributions induce a non-zero $\sigma_{xxx}$ and the Hall conductivities $\sigma_{yyx}=\sigma_{yxy}$ and $\sigma_{xyy}$. 
We emphasize that the magnetic field dependence of the NL resistivity (discussed below) originates primarily from the diagonal component of the NL conductivity, $\sigma_{xxx}$. This particular NL conductivity component is generally finite in a system where mirror symmetry is broken along the $x$-axis. 


From Fig.~\ref{fig_3} it is evident that all of the NL conductivities broadly follow two features--- i) all the NL conductivity contributions are peaked near the band edge and they decrease as we move away from the band edge after the initial rise and ii) the peaks are not located exactly at the band edge. 
The appearance of the peaks near the band edge can be understood from the fact that all these conductivity contributions  originate from the Berry curvature or the OMM which are primarily concentrated near the band edge. As we move away from these hotspots, the strength of the Berry curvature and the OMM decreases.
The fact that the peaks are not exactly at the band edges is related to the reason that NL conductivities originate from the combined effect of tilt and geometric quantities. The position of the peaks of the conductivities in the $\mu$-axis can be explicitly calculated from the analytical expressions provided in Table~\ref{table_1}.
If we consider $\sigma_{abc} \propto (a + br^2 +c r^4)/\mu^\nu$, then the peak position is found to be $\mu_0 = \Delta [(-(\nu+2)b \pm \sqrt{((\nu+2)^2b^2 -4\nu(\nu+4)ac})/(2\nu a)]^{1/2}$. 
Another interesting finding from the analytical results of $\sigma^{\rm OMM}$ [see the last column of Table~\ref{table_1}] is that unlike the other three of the contribution~(NAH, AL and B) the OMM contribution does not vanish exactly at $\Delta = \mu$. We note that unlike the $\sigma^{\rm NAH}$ and $\sigma^{\rm AL}$, the NL conductivities $\sigma^{\rm B}$ and $\sigma^{\rm OMM}$ changes sign as a function of $\mu$. This can be anticipated from analytical results which have two roots in $\mu$-axis.

\subsection{Nonlinear resistivity}

Having calculated the NL conductivities, we now turn our focus to the NL magnetoresistivity. For the tilted massive Dirac model of Eq.~\eqref{ham}, the Drude conductivity is calculated to be $\sigma_{xx}= g_s g_v\frac{e^2\tau_{\omega}}{\hbar^2}\frac{\mu}{4\pi} (1 - r^2)$. The linear classical Hall conductivity is calculated to be $\sigma^{\rm L}_{xy}=-\sigma^{\rm L}_{yx} = -g_s g_v\frac{e^3\tau_\omega^2 B}{\hbar^4}\frac{v_F^2}{4\pi} \left(1-r^2\right)$ and the OMM induced intrinsic Hall conductivity is obtained to be $\sigma^{\rm O}_{xy}=-\sigma^{\rm O}_{yx} = -g_s g_v\frac{e^3 B}{\hbar^2}\frac{v_F^2}{8\pi\mu^2} r^2$. Note that the linear conductivities presented above are calculated upto zeroth order in the tilt for simplicity, and this does not change our results qualitatively. 

To understand the dependence of the NL resistivity on the magnetic field to the lowest order, we use the magnetic field dependence of the different linear and NL conductivities in Eq.~\eqref{NL_resistance}, to obtain 
\be \label{NL_resistance_x}
\begin{pmatrix}
\tilde \rho_{xx}^{(2)} \\
\tilde \rho_{yx}^{(2)} 
\end{pmatrix}
\sim
\begin{pmatrix}
 B^0 & B^1  \\
B^1 & B^0 
\end{pmatrix}
\begin{pmatrix}
B^1& B^0 & B^0 & B^1 \\
B^0 & B^1 & B^1 & 0
\end{pmatrix}
\begin{pmatrix}
B^0\\
B^1\\
B^1\\
0
\end{pmatrix} ~.
\ee
Since we are only interested in the lowest order magnetic field dependence of the NL resistivity, we have neglected the quadratic $B$ dependence, and have thus put $\rho_{xy}^2 = 0$. Focussing on the lowest order magnetic field dependence, we find $\tilde \rho_{xx}^{(2)}\propto  B + {\cal O}(B^3)+\ldots$, and $\tilde \rho_{yx}^{(2)} \propto B^0 + {\cal O}(B^2)+\ldots$ . Interestingly, this magnetic field dependence of the second order resistivity is in contrast to the linear order resistivity, for which we have $\rho_{xx}^{(1)}\propto B^0$, while $\rho_{xy}^{(1)}\propto B$. 

In a more general form, valid for all 2D materials, we obtain
\bea  \label{r_xx_2}
\tilde \rho_{xx}^{(2)} &=& {-}\rho_{xx}^2[\rho_{xx} \sigma_{xxx} + 2\rho_{yx} \sigma_{xyx} + \rho_{xy} \sigma_{yxx}]~, 
\\ \label{r_yx_2}
\tilde \rho_{yx}^{(2)}  &=& {-} \rho_{xx}^2 \rho_{yy} \sigma_{yxx}~.
\eea
Equation~\eqref{r_xx_2} for the NL resistivity is calculated here for the first time (to the best of our knowledge), while Eq.~\eqref{r_yx_2} is the NL (anomalous) Hall resistivity, an experimental manifestation of the NL Hall conductivity predicted by Sodemann and Fu~\cite{Sodemann15}. From Eq.~\eqref{r_xx_2} it can be clearly seen that the magnetic field dependence of the NL resistivity comes from both the linear as well as NL conductivities. However, all the NL conductivities arise from the presence of a finite OMM and the Berry curvature. Thus, we conclude that the NL resistivity are induced by the quantum geometric properties of the electron wave-function, and they are of purely quantum mechanical origin.  



In the limit $\omega \tau \ll 1$ (generally valid for transport experiments), we can express the magnetoresistivity and Hall resistivity for the massive tilted Dirac Hamiltonian, in a simple form as
\bea \label{R_xx_2}
\tilde \rho_{xx}^{(2)} &=& {-} \frac{ 3\pi^2 \hbar^2 \Delta v_tv_F^2 r^2}{e^2 \tau \mu^6 (1-r^2)^3}  B~,
 \\ \label{R_yx_2}
\tilde \rho_{yx}^{(2)} &=& {-} \frac{3 \pi^2 \hbar^2 \Delta v_t}{2e^3\tau^2\mu^5(1-r^2)^2}~.
\eea
Here, we have neglected the contribution of the OMM induced intrinsic Hall conductivity in the linear response, which is relatively smaller than the classical Hall conductivity (see Appendix~\ref{exct_nn_lnr_res} for exact form).  
Clearly, the magnetoresistivity and Hall resistivity have different scattering time as well as chemical potential dependence. The $\Delta$ factor signifies the broken inversion symmetry, which enables finite values of the Berry curvature and the OMM. The factor $v_t$ highlights the effect of tilt, or the anisotropy of the Fermi surface or band velocity, which is pivotal to obtain NL conductivities as discussed earlier. 

\begin{figure}[t!]
\includegraphics[width=\linewidth]{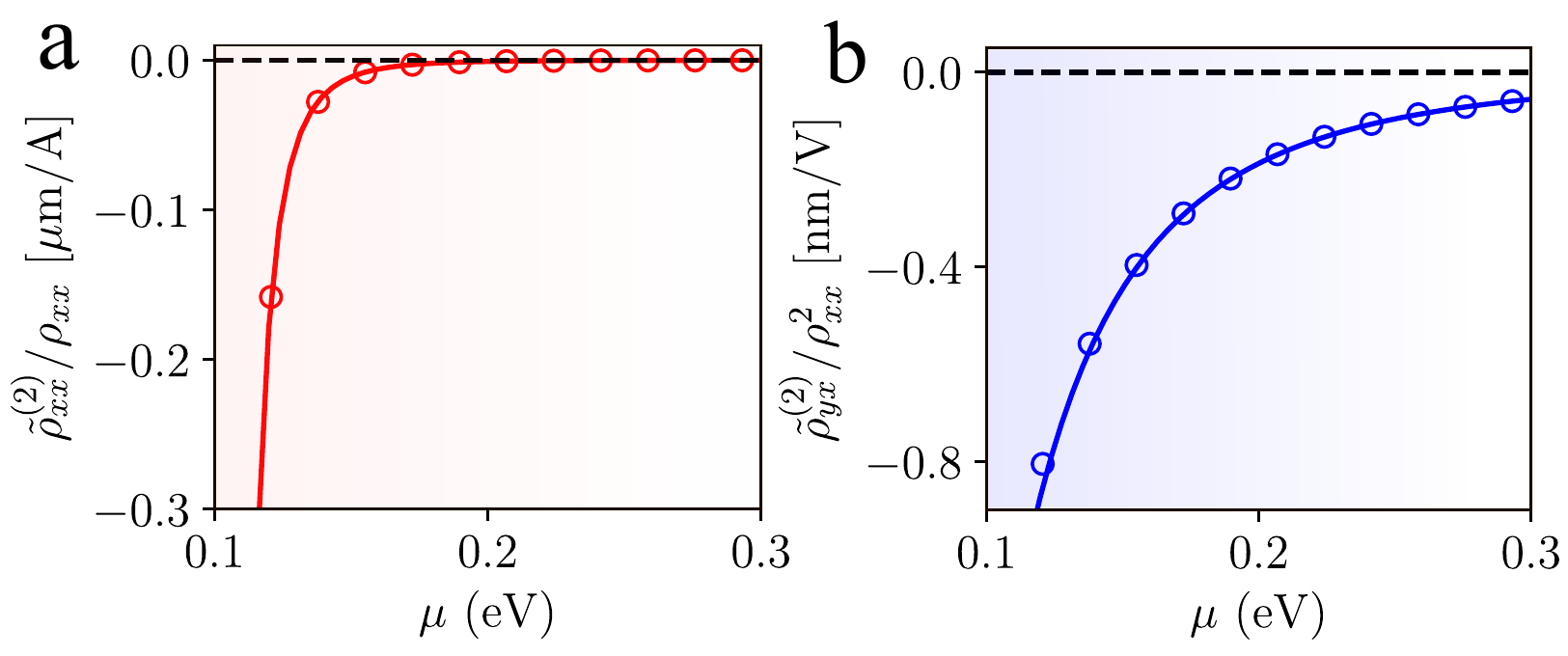}
\caption{(a) The variation of the scattering time independent a) NL resistivity ratio, $\tilde \rho_{xx}^{(2)}/\rho_{xx}$ and b) the NL Hall resistivity ratio, $\tilde \rho_{yx}^{(2)}/\rho_{xx}^2$, with the chemical potential $\mu$. The open circles correspond to the numerical result ($T=50$ K), and the solid-line represents our analytical result at zero temperature. The Hamiltonian parameters are same as those in Fig.~\ref{fig_2} and we have considered $\tau=1$ ps.}
\label{Fig4}
\end{figure}

From Eqs.~\eqref{R_xx_2}-\eqref{R_yx_2}, we define two experimentally relevant quantities which are independent of the scattering time scale. For the resistivity we define $\tilde \rho_{xx}^{(2)}/\rho_{xx}$ which is equivalent to the ratio of NL longitudinal voltage to the linear voltage (mutiplied by current)  [$V_{\parallel}^{(2)}/(V^{(1)}_{\parallel} I_x)$]. The variation of this quantity with the chemical potential is shown in Fig.~\ref{Fig4}(a). For the NL Hall resistivity, we define $\tilde \rho_{yx}^{(2)}/\rho_{xx}^2$ which is equivalent to the ratio of the NL Hall voltage to the square of the linear voltage [$V_{\perp}^{(2)}/(V_{\parallel}^{(1)})^2 $] and the corresponding chemical potential dependence is highlighted in Fig.~\ref{Fig4}(b). We find that the ratio of  $\tilde \rho_{xx}^{(2)}/\rho_{xx}$ is finite in a small region in vicinity of the band-edge,  while the ratio $\tilde \rho_{yx}^{(2)}/\rho_{xx}^2$ is finite in vicinity of the band-edge over a relatively larger region of $\mu$.

\section{Discussion}
\label{discussion}
In this paper, we have considered time reversal symmetric systems for two reasons. First, for TRS preserving systems the NL Drude conductivity vanishes, hence the predicted NL magnetoconductivities make the total NL resistivity finite. 
Second, the NL resistivity is caused solely by the geometric quantities, namely the Berry curvature and the OMM. These two observations make the experimental realization of the predicted NL resistivity and its physical origin very clear.
In contrast, in systems where both SIS and the TRS are broken, like doped magnetic semiconductors~\cite{bhalla_PRL2020_resonant}, the NL magnetoresistivity will be accompanied by an additional contribution from Drude resistivity and the classical Lorentz force contribution, as shown explicitly in Appendix~\ref{NDF_nonlinear}.
Interestingly, this novel NL magnetoresistivity of quantum mechanical origin, is not restricted to 2D systems only and it can also be finite for three dimensional systems~\cite{Golub20, Chan17, Cortijo16}.  

We emphasize that the relaxation time approximation, which we have used in our paper, while being very insightful, is a simplified approach. More rigorous approaches, developed in Refs.~[\onlinecite{Sekine18, Knoll19, Zhou19, Xiao20}] and others, can be used to include the field dependence in the scattering time. This remains as a future direction for us to explore NL magneto-resistivity after including the electric field dependence of the scattering timescale, along with skew-scattering. One can also extend our semiclassical approach to the quantum kinetic framework~\cite{sekine_PRB2017_quantum, sekine_PRB2020_quantum, huang_arxiv2021_resonant} with a magnetic field, though we believe that this is likely to produce results similar to what we have in this paper. Furthermore, 
going beyond the semiclassical regime which is valid for small magnetic fields, it will be interesting to see how our results change for large magnetic fields which gives rise to Landau levels. 


\section{Conclusion}
\label{conclusion}


To summarize, we explore the impact of band geometric quantities on the the second order NL magnetotransport in 2D anisotropic systems. Specifically, we study the second order NL magnetoresistivity, which relates the quadratic NL voltage generated in response to an applied current in crystalline materials. We show that the interplay of the Berry curvature, the OMM and the Lorentz force can induce NL resistivity in 2D systems, which is purely quantum mechanical in origin.
We find that in presence of a {\it perpendicular} magnetic field the NL magnetoreisitivity  has non-trivial linear-$B$ dependence. In intrinsically time reversal symmetric systems, where the Drude contribution to the longitudinal NL conductivity is identically zero, the predicted NL magnetoresistivity is the only finite NL resistivity contribution. 
Our findings pave the way for further understanding of the non-trivial transport signature of band geometry in quantum materials.

\section{acknowledgement}
We sincerely thank Sougata Mardanya and Pankaj Bhalla for useful discussions. We acknowledge  Science and Engineering Research Board (SERB) - India, and the Department of Science and Technology (DST) - India, and Indian Institute of technology Kanpur for financial support. 

\appendix
\section{Defining the nonlinear resistivity matrix}
\label{define_NL_R}
In this section of Appendix, we present a general approach of defining the NL resistivity matrix from the linear and NL conductivities.
The definition of resistance follows from the familiar Ohm's law. If voltage drop $V$ is measured in presence of constant current $I$, then the resistance $R$ is defined as the ratio of these two quantities, $R = \frac{V(I)}{I}$.
Within the linear response theory, voltage is considered to be linearly proportional to current and hence the resistance is current independent. However, this dependence of voltage on current can be nonlinear in general, which makes the resistance to be current dependent. Up to second order in current, the voltage can be expressed as~\cite{He19}
\be \label{voltage_expt}
V_i = V_i^{(1)} + V_i^{(2)} =   R_{ij}^{(1)}  I_j  +  \tilde R_{ij}^{(2)}  I_j^2~.
\ee
Here, the voltage $V^{(1)}(V^{(2)})$ is linear (quadratic) in current, $R_{ij}^{(1)}$ is the linear resistance and $\tilde R_{ij}^{(2)}$ represents NL resistance. Strictly speaking, $\tilde R_{ij}^{(2)}$ does not have the dimension of resistance, but we will still call it as NL resistance in spirit. In the linear response regime, one measures the voltage $V = V^{(1)}$ in response to an applied current $I$, and  resistance can be obtained using $R_{ij}^{(1)}=V/I$. From the theoretical point of view, instead of resistance we calculate the conductivity $\sigma$ from the relation $\bm j = \sigma \cdot \bm E$ where $\bm E$ and $\bm j$ are the applied electric field and generated current density, respectively. If we identify the measured voltage $V$ with $\bm E$ and the current $I$ with $\bm j$, then the resistance can be easily connected to the conductivity as $R^{(1)} \sim \rho = 1/\sigma$, $\rho$ being the linear resistivity. However, the scenario for $\tilde R_{ij}^{(2)}$ is not as simple as the linear resistivity and to remedy this we define the quantity $\tilde R_{ij}^{(2)}$ in terms of the NL conductivities below.

To do this, first we define the resistance as %
\be \label{Ohm_resis_curr}
R = \dfrac{V}{I(V)} \propto \dfrac{\bm E}{{\bm j}({\bm E})},
\ee
where the current varies as a function of the electric field. Beyond the linear response regime, the first NL correction in current density appears as ${\bm j}^{(2)} = \chi \cdot \bm E \cdot \bm E $. Note that in the main text we have used $\sigma_{abc}$ instead of $\chi$. Here we use $\chi$ to distinguish it from the linear conductivity $\sigma$. Including the NL electric field dependence of ${\bm j}$ upto second order, the resistance can be further expressed in the following way
\be
R \propto \dfrac{\bm E}{\sigma \cdot \bm E + \chi \cdot \bm E \cdot \bm E} = \dfrac{1}{\sigma} - \dfrac{1}{\sigma} \cdot \dfrac{\chi \cdot \bm E\cdot \bm E}{\bm j}~.
\ee
In the second term we have approximated $\sigma \cdot {\bm E} \approx {\bm j}$.
Identifying the measured current $I$ with $\bm j$ in the above equation, and multiplying both sides by $j$ we get
\be\label{resi}
V \propto  \rho j - \rho \cdot \dfrac{ \chi \cdot \bm E\cdot \bm E}{j^2}j^2~.
\ee
Note that $j^2$ is a scalar quantity (not a matrix) and we can compare this with Eq.~\eqref{voltage_expt}, as $j^2 =j_x^2 + j_y^2 +j_z^2 $. Comparing Eq.~\eqref{resi} with Eq.~\eqref{voltage_expt} we can identify
\be \label{Ohm_resis_nn_lnr}
\rho^{(1)} = \rho ~~\mbox{and}~~~ \tilde \rho^{(2)}  = - \rho \cdot \dfrac{ \chi \cdot {\bm E} \cdot {\bm E}}{j^2}~.
\ee
Now converting the electric field (voltages) to current as $\bm E = \rho \cdot \bm j $ in the second equation, we obtain the explicit form of $\tilde \rho^{(2)}$ as
\be \label{NL_resis}
\tilde \rho^{(2)}  = {-}\rho \cdot \chi \cdot \rho \cdot \rho~.
\ee
%
Equation~\eqref{NL_resis} is the general expression of NL resistivity matrix in terms of the NL conductivity and linear resistivity.
We have used this equation to calculate Eq.~\eqref{NL_resistance} of the main text.

\section{Linear conductivities}\label{NDF_linear}

In this section of Appendix, we provide the detailed calculation of the NDF upto linear order in ${\bm E}$-field and linear order in ${\bm B}$-field. For this we consider the ansatz $\delta g_1(t) = f_1^\omega e^{i\omega t} + f_1^{\omega *} e^{-i\omega t}$. Putting this in Eq.~\eqref{boltzmann}, we get
\begin{equation} \label{dlta_n_1}
f_1^\omega  = \sum_\nu \left(D \tau_{\omega}\hat L_{\rm B}\right)^\nu \left(D\dfrac{e \tau_\omega }{\hbar}  {\bm E} \cdot \nabla_{\bm k} \tilde f \right)~.
\end{equation}
Considering low strength of magnetic field, we expand the series in Eq.~\eqref{dlta_n_1} to get various orders of magnetic field dependence. Separating the magnetic field dependences, we write the NDF as $f_1^\omega = f_{10}^{\omega} + f_{11}^{\omega}$, where the first subscript denotes the order of electric field and the second subscript denotes the order of magnetic field. This approach of expansion of the $L_{\rm B}$ operator is very common in the textbooks and is known as Zener-Jones method~\cite{Hurd72, Gao19}. The scattering time independent~\cite{Gao19} equilibrium part of the NDF, after the Taylor expansion, is given by
\be
\tilde f  = f - \epsilon_{\rm m}  f^\prime~.
\ee
The NDF proportional to the linear order of scattering time is given by
\bea \label{dlta_n_1_0}
f^{\omega }_{10}  &=& e \tau_\omega  {\bm E} \cdot  {\bm v} f^\prime~,
\\ \label{dlta_n_1_1}
f^{\omega}_{11}(\gamma, \xi)  &=& -e \tau_\omega  
 \left[ \gamma \Omega_{\rm B}  {\bm v}
 f^\prime 
+\xi \left({\bm v}_{\rm m} f^\prime + \epsilon_{\rm m} {\bm v} f^{\prime \prime} \right)\right]\cdot  {\bm E}.~~~~~
\eea
Equation~\eqref{dlta_n_1_0} is the magnetic field independent part from which the linear Drude conductivity originates. To point out the origin of magnetic field dependences in NDF, we use coefficient $\gamma$ for the phase-space factor and $\xi$ for the OMM. The quadratic scattering time dependent part of the NDF is given by
\bea\label{f_loretz_1}
f^{\omega}_{11} (\alpha) &=& \alpha e \tau_\omega^2   \hat L   {\bm v}\cdot  {\bm E} f^\prime~.
\eea
Here, we have defined $ \hat L \equiv \frac{e }{\hbar}({\bm v} \times {\bm B})\cdot {\bm \nabla}_{\bm k}$ and $\alpha$ denotes the origin of magnetic field dependence from the Lorentz force.  

The magnetic field independent current of fundamental frequency can be written as ${\bm j}_{10}(t) = {\bm j}_{10}^\omega e^{i \omega t} + {\bm j}_{10}^{\omega*} e^{-i \omega t} $. From this we define the magnetic field independent conductivities as
\bea \label{sigma_A}
\sigma_{ab}(\propto \tau^0) = -\dfrac{e^2}{\hbar} \varepsilon_{abd} \int[{d \bm k}]\Omega_d ~f ,
\\\label{sigma_D}
\sigma_{ab}(\propto \tau) =- e^2\tau_\omega \int[d\mathbf{k}] v_a v_b ~f^\prime~.
\eea
Equation ~\eqref{sigma_A} is the anomalous Hall conductivity which vanishes in presence of TRS. Equation~\eqref{sigma_D} is the ordinary Drude conductivity. 

Similarly, the magnetic field dependent current in fundamental frequency can be written as ${\bm j}_{11}(t) = {\bm j}_{11}^\omega e^{i \omega t} + {\bm j}_{11}^{\omega*} e^{-i \omega t}$.
The corresponding conductivities, after separating various scattering time dependences, can be written as
\bea
\sigma_{ab}(\propto\tau^0)&=& \xi \dfrac{e^2}{\hbar}\varepsilon_{abd}\int [d{\bm k}]\Omega_d\epsilon_{\rm m} f^\prime ,
\\\nn
\sigma _{ab}(\propto\tau^2) &=&  -\alpha  e^2 \tau_\omega^2  \dfrac{eB}{\hbar}\int[d{\bm k}] v_a (v_y \partial_{k_x} v_b -v_x \partial_{k_y} v_b)  f^\prime,
\\
\eea
and 
\bea \nn
\sigma_{ab}(\propto\tau)&=& e^2 \tau_\omega  \int  [d{\bm k}]  \Big[ \xi v_{{\rm m}a}v_b f^\prime 
 + \gamma {v_a}\Omega_{\rm B}v_bf^\prime~~~~~~~~~
\\
&&~~~~~~~+ \xi v_a\left(v_{{\rm m}b} f^\prime + \epsilon_{\rm m} v_b f^{\prime \prime} \right) \Big]~.
\eea
In presence of TRS (broken SIS), $(\epsilon_{\rm m}, {\bm \Omega})(-{\bm k})= -(\epsilon_{\rm m}, {\bm \Omega})({\bm k})$, ${\bm v}(-{\bm k}) = -{\bm v}({\bm k})$ and ${\bm v}_{\rm m}(-{\bm k}) = {\bm v}_{\rm m}({\bm k})$, hence all the conductivities $\propto \tau$ vanish. The only terms that survive are a) the Lorentz force contribution $\propto \tau^2$ which gives rise to the classical Hall effect and b) the anomalous velocity and OMM contribution that is $\propto \tau^0$. We have written these non-zero contributions in Eqs.~\eqref{lorentz}-\eqref{OMM_linear} of the main text.

\section{Nonlinear conductivities}\label{NDF_nonlinear}

In this section of Appendix, we calculate the NDF quadratic in ${\bm E}$-field and linear in ${\bm B}$-field. The corresponding master equation for the rectification part of the NDF is given by 
\begin{equation}\label{dlta_n_2_dc}
f_2^0 =\sum_\nu \left(\alpha D \tau \hat L_{\rm B}\right)^\nu D\frac{e\tau}{\hbar} {\bm E}^* \cdot {\bm \nabla}_{\bm k} f_1^{\omega} .
\end{equation}
It is straight forward to calculate the rectification part of the NDF from Eq.~\eqref{dlta_n_2_dc}. However, below we provide the second harmonic NDF obtained from Eq.~\eqref{dlta_n_2}. We expand Eq.~\eqref{dlta_n_2} in orders of magnetic field and separate the NL NDF as $f_2^{2\omega}=f^{2\omega}_{20} +f^{2\omega}_{21}$. The magnetic field independent NL NDF is given by
\be\label{dlta_n_2_0}
f^{2\omega}_{20} = \frac{e^2\tau_{2\omega}\tau_\omega}{\hbar} {\bm E}\cdot {\bm \nabla}_{\bm k} \left( {\bm E} \cdot  {\bm v} f^\prime\right).
\ee
This coupled with band gradient velocity generates the NL Drude conductivity. The magnetic field dependent NL NDF proportional to the square of scattering time is given by
\bea
&f^{2\omega}_{21}(\gamma, \xi) =  - \dfrac{e^2\tau_{2\omega}\tau_\omega}{\hbar} {\bm E}  \cdot \left[\gamma \Omega_{\rm B} {\bm \nabla}_{\bm k} ( {\bm E}\cdot {\bm v}f^\prime)  \right.
\\[1ex]\nn
&  \left.  \gamma {\bm \nabla}_{\bm k} (\Omega_{\rm B}{\bm E}\cdot {\bm v}f^\prime) + \xi  {\bm \nabla}_{\bm k} \{ {\bm E}\cdot({\bm v_{\rm m}}f^\prime + \epsilon_{\rm m}{\bm v}f^{\prime\prime})\} \right].
\eea 
The NL NDF proportional to the cubic order of scattering time is calculated to be
\bea  \nn
f^{2\omega}_{21} = \frac{e^2\tau_{2\omega}\tau_\omega}{\hbar} \left[\tau_{2\omega}\hat{L} ( {\bm E} \cdot {\bm \nabla}_{\bm k}( {\bm E}\cdot {\bm v}f^\prime))~~~~~~~~  \right.
\\
\left.~ + \tau_{\omega}{\bm E}  \cdot {\bm \nabla}_{\bm k}( \hat{L}{\bm E}\cdot {\bm v}f^\prime) \right].
\eea
%
Using these expressions of NDF we now calculate the general expressions of NL conductivities. Note that the NL responses can either be generated via coupling of anomalous velocity with linear NDF or through coupling of band velocity of Bloch electrons with the NL NDF. 

The magnetic field independent second harmonic NL current can be written as ${\bm j}_{20}(t) = {\bm j}^{2\omega}_{20}e^{i2\omega t} +{\bm j}^{2\omega*}_{20}e^{-i2\omega t} $ and the corresponding NL conductivities are yield to be
\bea \label{j_2_ano}
\sigma_{abc} (\propto \tau) &=& - \dfrac{e^3 \tau_\omega}{2 \hbar} \varepsilon_{abd} \int[d{\bm k}]\Omega_dv_c ~f^\prime ~,
\\ \label{j_2_Drude}
\sigma_{abc} (\propto \tau^2) &=& -\dfrac{ e^3 \tau_\omega \tau_{2\omega} }{2 \hbar}\int [d{\bm k}] v_a \partial_{k_b}v_c f^\prime.~~~
\eea
%
In presence of TRS, the NL conductivity $\propto \tau$ [Eq.~\eqref{j_2_ano}] survives, while in presence of SIS both Eqs.~\eqref{j_2_ano}-\eqref{j_2_Drude} vanish identically. So unlike the linear Drude conductivity which is always non-zero, the NL Drude conductivity, Eq.~\eqref{j_2_Drude}, vanishes in presence of any of the symmetries (among TRS and SIS). In systems where both the symmetries are absent, the NL Drude conductivity may give rise to bilinear magnetoresistance if one considers the Zeeman coupling~\cite{Zhang18, Dyrdal20}. This theory has been used to explain some recent experimental observations in the topological insulator surface states~\cite{He18, He19}.

The magnetic field dependent second harmonic NL current can be expressed as ${\bm j}_{21}(t) = {\bm j}^{2\omega}_{21}e^{i2\omega t} +{\bm j}^{2\omega*}_{21}e^{-i2\omega t} $. The NL responses are induced by the OMM velocity, anomalous velocity and the band gradient velocity. 
Below, we will express the NL conductivities as $\sigma_{abc} = \frac{e^3}{\hbar} \int [d{\bm k}] \tilde \sigma_{abc}$ for compactness.  The anomalous velocity induced NL conductivity a) $\propto \tau$ is given by 
\be 
\tilde  \sigma_{abc} (\propto \tau)  =  \tau_\omega \varepsilon_{abd}\Omega_d \left[
 \gamma \Omega_{\rm B} v_c
f^\prime 
 + \xi \left(v_{{\rm m}c} f^\prime + \epsilon_{\rm m} v_c f^{\prime \prime} \right) \right],
\ee
and b) $\propto \tau^2$ is given by
\be
\tilde \sigma_{abc} (\propto \tau^2)  =  - \tau_\omega^2 \varepsilon_{abd} \Omega_d \dfrac{eB}{\hbar} (v_y \partial_{k_x}v_c - v_x \partial_{k_y} v_c) f^\prime.
\ee
The OMM velocity induced NL conductivity is given by
\be 
\tilde \sigma_{abc} (\propto \tau^2)  =   \tau_\omega \tau_{2\omega}  v_{{\rm m}a}\partial_{k_b} v_c f^\prime .
\ee
Finally, the band gradient velocity induced conductivity a) $\propto \tau^2$ is given by
\bea \nn
\tilde \sigma_{abc} (\propto \tau^2) =    \tau_\omega \tau_{2\omega} v_a
 \Big[  \Omega_{\rm B}\partial_{k_b}v_cf'~~~~~~~~~~~~~~
\\
+   \partial_{k_b}\left \{ \Omega_{\rm B}v_cf'+ (v_{{\rm m}c} f' +\epsilon_{\rm m} v_c f'' ) \right \}
\Big],
\eea
and b) $\propto \tau^3$ is given by
\bea \nn \label{Lorentz}
\tilde \sigma_{abc} (\propto \tau^3)&=&-\tau_{2\omega}\tau_{\omega}  v_a \dfrac{eB}{\hbar} \Big[ \tau_{2 \omega } \big(v_y\partial_ {k_x} 
 - v_x\partial_{k_y}\big)\partial_{k_b}v_c ~~~~
 \\
&&~~~~~~+ \tau_{\omega} \partial_{k_b} \big(v_y\partial_{k_x} -v_x\partial_{k_y}\big)v_c \Big] f^\prime.
\eea
We find that in presence of TRS, NL conductivities $\propto \tau$ and $\propto \tau^3$ vanish identically. Therefore NL conductivities that are $\propto \tau^2$ survive and are highlighted in the main text in Eqs.~\eqref{sigma_NAH}-\eqref{sigma_B}. We emphasize that since the indices $b$ and $c$ are dummy, the expressions of NL conductivities have to be symmetrized. To facilitate this symmetry we have wriiten the NL conductivities as $\sigma_{abc} = \sigma_{acb} = \frac{e^3}{\hbar}\int [d{\bm k}]\left[\tilde\sigma_{abc} + \tilde\sigma_{acb}\right]/2$ in the main text. It is important to note that in absence of both the SIS and TRS the NL conductivities $\propto \tau$ and $\propto \tau^3$ are expected to be non-zero and in that case NL resistivity can originate from the classical Drude and the Lorentz force effect. 

\section{Exact analytical expression of nonlinear resistivity}
\label{exct_nn_lnr_res}

In this section of Appendix, we provide the exact analytical expression of the NL resistivity including the effect of  OMM induced intrinsic Hall conductivity. We calculate the linear resistivities using the expressions: $\rho_{xx}=1/\sigma_{xx}$ and $\rho_{xy}=-\rho_{yx} = -\sigma_{xy}/\sigma_{xx}^2$. Using these expressions in Eq.~\eqref{r_xx_2} and considering $\omega \tau \ll 1$, we calculate
\be \label{exact}
\tilde \rho_{xx}^{(2)} = {-}\frac{3 \pi^2 \hbar^2  \Delta v_t v_F^2  r^2}{2 e^2\tau \mu^6(1-r^2)^3}\left[2 + \frac{\hbar^2}{\tau^2 \mu^2}\right]B~.
\ee
Here, the second term in the parentheses originates from the OMM induced intrinsic Hall effect. It is evident from this expression that the  OMM induced intrinsic Hall effect has a distinct signature in the NL resistivity. However, since in this paper we have considered $\hbar/\tau \ll \mu$, so we neglect it in the main text. Another interesting feature of Eq.~\eqref{exact} is that the NL resistivity diverges as we move towards $r \to 1$. This arises due to the fact that both the linear and NL conductivities vanish in this regime.

\bibliography{NL_MT_2D.bib}
\end{document}